\documentclass[aps,pre,preprint,groupedaddress,showpacs]{revtex4-1}
\usepackage{amsmath}
\usepackage[T1]{fontenc}
\usepackage[utf8]{inputenc}
\usepackage[pdftex]{graphicx}
\usepackage{xspace}
\usepackage{mathrsfs}
\newcommand{\km}{\ensuremath{\text{KM}\xspace}}
\newcommand{\nap}{\ensuremath{\text{Nap}\xspace}}
\newcommand{\kd}{\ensuremath{\text{Kd}\xspace}}
\newcommand{\mkm}{\ensuremath{m_{\km}}\xspace}

\newcommand{\ikd}{\ensuremath{I_{\kd}}\xspace}
\newcommand{\ikm}{\ensuremath{I_{\km}}\xspace}
\newcommand{\ina}{\ensuremath{I_{\nap}}\xspace}
\newcommand{\iext}{\ensuremath{I_{\text{ext}}}\xspace}

\newcommand{\pcmsq}{\ensuremath{\text{cm}^{-2}}\xspace}
\newcommand{\msiemens}{\ensuremath{\text{mS}}\xspace}
\newcommand{\farad}{\ensuremath{\text{F}}\xspace}
\newcommand{\mvolt}{\ensuremath{\text{mV}}\xspace}
\newcommand{\msec}{\ensuremath{\text{ms}}\xspace}
\newcommand{\uamp}{\ensuremath{\mu\text{A}}\xspace}
\newcommand{\ttof}[1]{\ensuremath{\mathscr{T}\left \{ #1 \right\}}\xspace}
\newcommand{\tof}[1]{\ensuremath{\mathcal{T}\left \{ #1 \right \}}\xspace}

\newcommand{\stg}{\textsc{stg}\xspace}
\newcommand{\cpg}{\textsc{cpg}\xspace}
\newcommand{\isi}{\textsc{isi}\xspace}

\begin{document}
\title{Noise, transient dynamics, and the generation of realistic
interspike interval variation in square-wave burster neurons}

\author{Bóris Marin}
\email[]{bmarin@if.usp.br}
\affiliation{Instituto de Física, Universidade de São Paulo, Brazil}
\altaffiliation[Present address: ]{Department of Neuroscience, Physiology
and Pharmacology, University College London, London, UK}

\author{Reynaldo Daniel Pinto}
\email[]{reynaldo@ifsc.usp.br}
\affiliation{Instituto de Física de São Carlos, Universidade de São Paulo,
Brazil}

\author{Robert C. Elson}
\email[]{RobertElson@pointloma.edu}
\affiliation{Institute for Nonlinear Science, University of California, San Diego, CA 92093-0402, USA}
\altaffiliation[Present address: ]{Department of Biology, Point Loma Nazarene University, San Diego, CA 
92106, USA}

\author{Eduardo Colli}
\email[]{colli@ime.usp.br}
\affiliation{Instituto de Matemática e Estatística, Universidade de São Paulo,
Brazil}


\date{\today}

\begin{abstract}
  
First return maps of interspike intervals for biological neurons that
generate repetitive bursts of impulses can display stereotyped
structures (neuronal signatures). Such structures have been linked to
the possibility of multicoding and multifunctionality in neural
networks that produce and control rhythmical motor patterns. In some
cases, isolating the neurons from their synaptic network reveals
irregular, complex signatures that have been regarded as evidence of
intrinsic, chaotic behavior.

We show that incorporation of dynamical noise into minimal neuron
models of square-wave bursting (either conductance-based or abstract)
produces signatures akin to those observed in biological examples,
without the need for fine-tuning of parameters or ad hoc constructions
for inducing chaotic activity. The form of the stochastic term is not
strongly constrained, and can approximate several possible sources of
noise, e.g. random channel gating or synaptic bombardment.

The cornerstone of this signature generation mechanism is the rich,
transient, but deterministic dynamics inherent in the square-wave
(saddle-node/homoclinic) mode of neuronal bursting. We show that noise
causes the dynamics to populate a complex transient scaffolding or
skeleton in state space, even for models that (without added noise)
generate only periodic activity (whether in bursting or tonic spiking
mode).

\end{abstract}


\maketitle

\section{Introduction}

First return maps of interspike intervals (\isi{}s) of bursting
biological neurons reveal characteristic patterns of firing
sequences~\cite{Szucs2003, Szucs2005, Zeck2007, Garcia2005}. In
invertebrate central pattern generators (\cpg{}s)~\cite{Szucs2003,
Szucs2005}, \isi return maps consist of a specific arrangement of
clusters, called a neuronal signature~\cite{Szucs2003, Szucs2005}.
The reproducibility of these signatures allows the identification of
neuronal types in circuits presenting different bursting frequencies,
duty cycles and number of spikes per burst, even across biological
species~\cite{Brochini2011}. Moreover, the signature reflects circuit
connectivity~\cite{Szucs2003, Szucs2005}, information in synaptic
input patterns~\cite{Brochini2011} and the modulation of network
operation~\cite{Zhurov2006}.

In realistic electrophysiological neuronal models, several dynamical
variables and parameters interact in nonlinear ways to produce complex
activity patterns, such as quiescence, tonic spiking and bursting. The
spiking-bursting activity may be periodic or chaotic. A burst of
spikes, taken as a whole, might function as a robust unit of neural
information~\citep{Lisman1997, Doiron2003, Izhikevich2003,
Oswald2004}. In contrast, the possibility of information coding
\emph{within} bursts has received little attention.

The information-processing properties of \cpg{}s have been explored in
model circuits inspired by the networks in the crustacean
stomatogastric ganglion (\stg)~\cite{DeBorjaRodriguez2002, Latorre2006,
Tristan2004}. The authors proposed a \cpg that generates a steady
rhythm of bursts but also responds to or recognizes the signatures
produced by its individual neurons. Analysis in the \stg has shown
that there is a neuron-to-neuron flow of information within a
bursting, rhythm-generating network~\cite{Brochini2011}.

The \isi return maps of many \stg neurons change considerably when the
cells are disconnected from their synaptic circuit. In these isolated
neurons, the \isi sequences \emph{within} each burst vary
\emph{between} bursts, this variation growing exponentially as bursts
evolve and the spike train progresses~\cite{Elson1999}. This activity
has been classified as chaotic bursting~\cite{Selverston2000}. The
neurophysiological mechanisms of this behavior have remained
elusive. Detailed, conductance-based models generally produce regular
activity when parameters are set to biologically plausible values
~\cite{Prinz2003}. Deterministic neuron models operating in chaotic
regimes can generate irregular (non-periodic, broad spectrum)
time-series, but their \isi return maps are highly structured, because
chaotic trajectories are confined to particular regions of state
space~\cite{Guckenheimer2002, Shilnikov2005, Terman1991,
Medvedev2006}. Moreover, the production of chaotic activity involves
fine-tuning of model parameters in order to meet strict criteria,
e.g. being close to spike-adding bifurcations~\cite{Terman1991}.

In contrast, state-space trajectories generated by stochastic
processes are not confined in this way, because noise is able to nudge
a dynamical system to populate the transient scaffolding (skeleton)
inherent in the dynamics. Accordingly, we here propose a mechanism to
generate the \isi signature of irregularly bursting neurons based on
the interplay of deterministic and stochastic dynamics. The noiseless
system does not need to be tuned to a chaotic regime, nor is there a
restrictive definition of the origin of the noise. We anticipate that
this approach can be applied to other problems as well, such as burst
alignment algorithms and noise-level estimation.

\section{Methods}

\subsection{Biological neurons: recordings and analysis}

The stomatogastric nervous system was removed from spiny lobsters,
\emph{Panulirus interruptus}, and pinned out in vitro in standard
\emph{Panulirus} saline~\cite{Elson1999}. The \stg, which contains the
rhythm-generating pyloric circuit, remained connected to anterior
ganglia whose descending modulatory influence sustain cellular
bursting activity. The lateral pyloric \textsc{lp} or a pyloric
dilator neuron \textsc{pd} neuron were disconnected from synaptic
input from other pyloric circuit neurons by photoinactivating or
deeply hyperpolarizing some presynaptic neurons and blocking inputs
from others pharmacologically~\cite{Miller1982, Bal1988}. After
synaptic isolation, neurons were impaled by two microelectrodes for
independent current injection and membrane potential recording.

Signature maps were obtained by detecting spikes via crossing a
threshold of $-35\text{mV}$. After each such crossing, new spike
detection was allowed only after the membrane potential timeseries
crossed a reset threshold of $-38\text{mV}$.

\subsection{Neuron Model}

Our analysis made use of a tridimensional conductance-based neuronal
model (\ref{eq:model}). This model has been introduced as a minimal
model for square-wave bursting~\cite{Izhikevich2000} and has been
previously analyzed in~\cite{Hitczenko2007, Medvedev2006}. It consists
of a two-dimensional fast subsystem coupled to a one-dimensional
slower one. The fast subsystem consists of a persistent sodium
current with instantaneous activation \ina,  and a potassium current
\ikd. The slow subsystem comprises the gating dynamic of an
\emph{M-type} potassium current \ikm. The biophysical parameters are
listed in table \ref{tab:pars}.

\begin{equation}
\begin{split}
  \label{eq:model}
  C\dot{V} & = \iext - g_{\text{leak}} (V - V_{\text{leak}}) 
  - \overbrace{\overline{g}_{\nap}m^{\infty}_\nap\left ( V - V_{\nap} \right)}^{\ina}
  - \underbrace{\overline{g}_i m_i \left ( V - V_\text{K} \right )}_{\text{I}_i}  \\
  \dot{m_i} & = \frac{m_i^\infty - m_i}{\tau_i} \quad i = [\kd, \km] \\
\end{split} 
\end{equation}

\begin{table}[h]
\caption{\label{tab:pars} Parameters for the currents in the
deterministic simulations (equation \ref{eq:model}). $V_i$ are the
ionic reversal potentials, $\bar{g}_i$ are maximum conductance
densities, $\tau_i$ are the timescales for each conductance. The
steady state activation functions are defined as $m^\infty_i = \{1 +
\exp{[(V^{1/2}_i-V)/k_i}]\}^{-1}$, where $V$ stands for the membrane
potential. Other passive parameters are the membrane specific
capacitance $C = 1\ \mu\farad\pcmsq$ and a \textsc{dc} bias $\iext =
5\ \uamp\pcmsq$.}
\begin{ruledtabular}
  \begin{tabular}{ccccc}
                               & \nap  & \kd   & \km   & leak   \\ \hline
  $V_i$ (\mvolt)               & 60    & -90   & -90   & -80    \\
  $\bar{g}_i$ (\msiemens\pcmsq)& 20    &  9    & 5     & 8      \\
  $V^{1/2}_i(\mvolt)$          & -19.9 & -25   & -21.2 & ---    \\
  $k_i (\mvolt)$               & 15    & 5     & 5     & ---    \\
  $\tau_i (\text{ms})$    & ---   & 0.152 & 20    & ---    \\
\end{tabular}
\end{ruledtabular} 
\end{table}

\begin{figure}[h]
\includegraphics[scale=1]{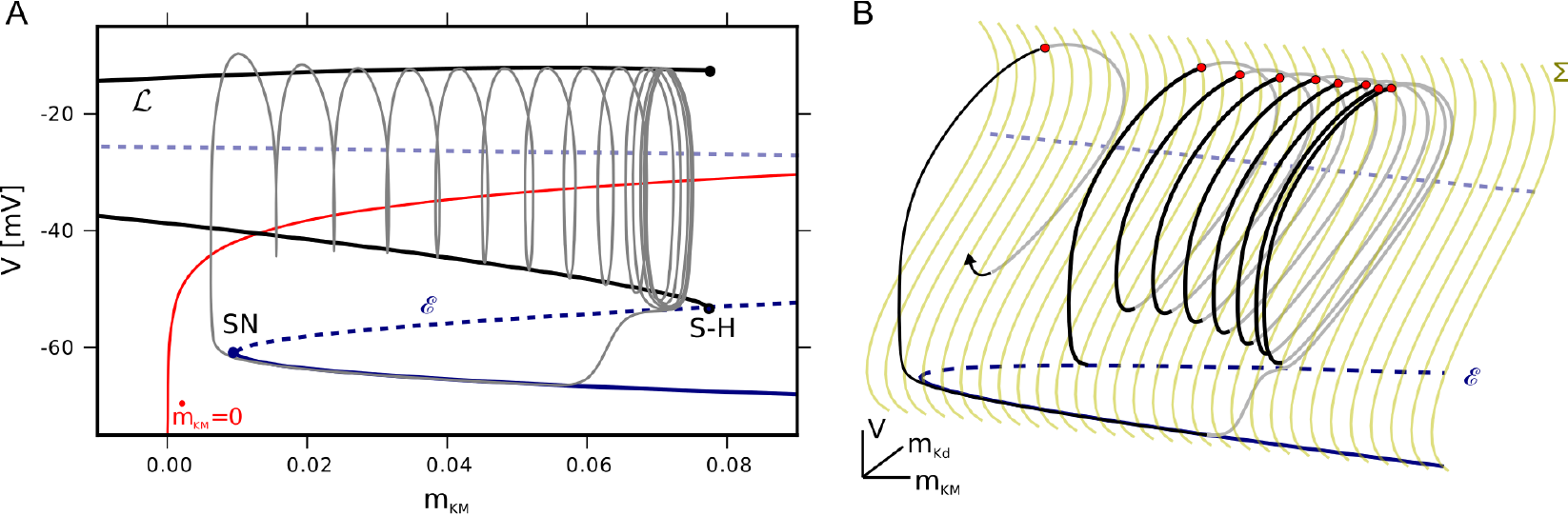}
\caption{\label{fig:phase_portrait}A: Phase portrait for the noiseless
neuronal model (eq. \ref{eq:model}). A noiseless bursting trajectory
is depicted in gray. The blue curves correspond to the fast subsystem
equilibrium branch $\mathcal{E}$, where solid/dashed lines denote
stable/unstable(saddle) states. The black ``tube'' depicts the spiking
manifold $\mathcal{L}$. The red curve is the \mkm nullcline:
trajectories below(above) it move towards smaller(larger) \mkm
values. Labelled points indicate bifurcations in the fast subsystem:
\textsc{sn}: saddle-node and \textsc{s-h}: saddle -- homoclinic
orbit. B: Full-model trajectory (black / gray curve) emanating from initial
condition in the unstable manifold of the saddle branch $\mathcal{E}$.
The Poincaré surface of section $\Sigma$ is depicted in beige. Red
points represent $\Sigma$ crossings, used in the reduced model
analysis.}
\end{figure}

Bursting activity is generated via an hysteretic loop. It can be
easily analyzed by considering the slow variable \mkm as a bifurcation
parameter ~\cite{Izhikevich2000, Fenichel1979}, which drives the fast
subsystem cyclically from a branch of equilibria (henceforth denoted
as $\mathcal{E}$) to a limit cycle manifold $\mathcal{L}$
(Fig. \ref{fig:phase_portrait}). When the trajectory (gray curve in
Fig. \ref{fig:phase_portrait}) slides along the stable equilibrium
part of $\mathcal{E}$ (solid segment of blue line marked as
$\mathcal{E}$), the full tridimensional system is in the
hyperpolarized, interburst phase. Since the flux is evolving below the
\mkm nullcline (red curve), it moves toward smaller \mkm values. The
stable equilibrium eventually loses stability in a saddle-node
bifurcation (point labeled \textsc{sn}), so the trajectory moves
towards the spiking manifold $\mathcal{L}$ (black ``tube''). Crossing
the \mkm nullcline leads the trajectories to move towards larger \mkm
values.  Spikes in the active phase of bursting correspond to full
revolutions around $\mathcal{L}$. This manifold disappears in a
saddle-homoclinic orbit bifurcation (point labelled \textsc{s-h}),
when it collides with the middle (saddle) segment of $\mathcal{E}$.

Several conductance-based models can give rise to square-wave bursting,
including those built to study systems as diverse as pancreatic
$\beta$--cells ~\cite{Chay1983, Terman1991}, neurons in the
pre-Bötzinger complex of the brain stem ~\cite{Butera1999, Best2005}
or hippocampal \textsc{ca1} pyramidal cells~\cite{Golomb2006}.  Our
analysis can be applied to any of these or even other systems,
provided that bursting involves a saddle middle branch in the
equilibrium curve of the fast subsystem.

As additional examples of square-wave bursting neuron signatures, we have
included those generated by the Hindmarsh-Rose three dimensional model
~\cite{Hindmarsh1984}, with parameters as in table \ref{tab:hr},
\begin{table}[h]
\caption{ \label{tab:hr}Parameter values for the Hindmarsh-Rose neuron
model, periodic bursting mode.}
\begin{ruledtabular}
\begin{tabular}{cccccccc}
  $a$ & $b$ & $c$ & $d$ & $s$ & $x_1$ & $r$ & $I$ \\\hline
  1 & 2.7 & 1 & 5 & 4 & -1.6 & 0.01 & 4
\end{tabular}
\end{ruledtabular}
\end{table}
and a model for neurons in the pre-Bötzinger complex (model and
parameters described in~\cite{Butera1999}, model 1). In both models,
the chosen parameter set supported periodic bursting activity.

\subsection{Stochastic dynamics}

Stochastic ion channel gating has been suggested to be the major
source of noise in isolated neurons~\cite{Rowat2004}. Since our
derivation of the generative model for \isi map signatures does not
impose constraints nor require a particular noise mechanism, we chose
to model stochastic gating using three different
approaches~\cite{Goldwyn2011}. For the \nap-\kd-\km{} model, we used a
Langevin approximation to microscopic gating schemes derived
in~\cite{Fox1997}. In this approximation, the subunit gating dynamics
$\dot{m_i}$ are complemented with a state dependent (multiplicative)
random forcing $\xi$, with zero mean and variance inversely
proportional to the number of channels $N$ in the membrane patch,
according to equation \ref{eq:langevin}.

\begin{equation}
\label{eq:langevin}
\begin{split}
&\dot{m_i} = \frac{m^{\infty}_i - m_i}{\tau_i} +  \xi_i \qquad 
\langle \xi_i\rangle = 0\\
&\langle \xi_i(t)\xi_i(t')\rangle =  \frac{m_i^\infty(1 -
m_i^\infty)}{N_i\tau_i}\delta(t-t') 
\end{split}
\end{equation}

For the Hindmarsh-Rose model, we opted for the current noise approach
~\cite{Goldwyn2011}, adding a stochastic force directly to the membrane
potential equation. Finally, for the pre-Bötzinger neuron
model, we used conductance noise~\cite{Goldwyn2011, Fox1997}, where the
stochastic terms are added to the conductance terms in the voltage
dynamics: $I_i = \overline{g}_i(m_i + \xi_i)(V -V_i)$.

The resulting stochastic differential equations were integrated numerically,
using the Euler-Maruyama scheme~\cite{Kloeden1992} with a fixed timestep of
$0.001\ \msec$.

\section{Results}

\subsection{ISI maps of irregular bursting in biological neurons}

\begin{figure}[htp]
\includegraphics[scale=1]{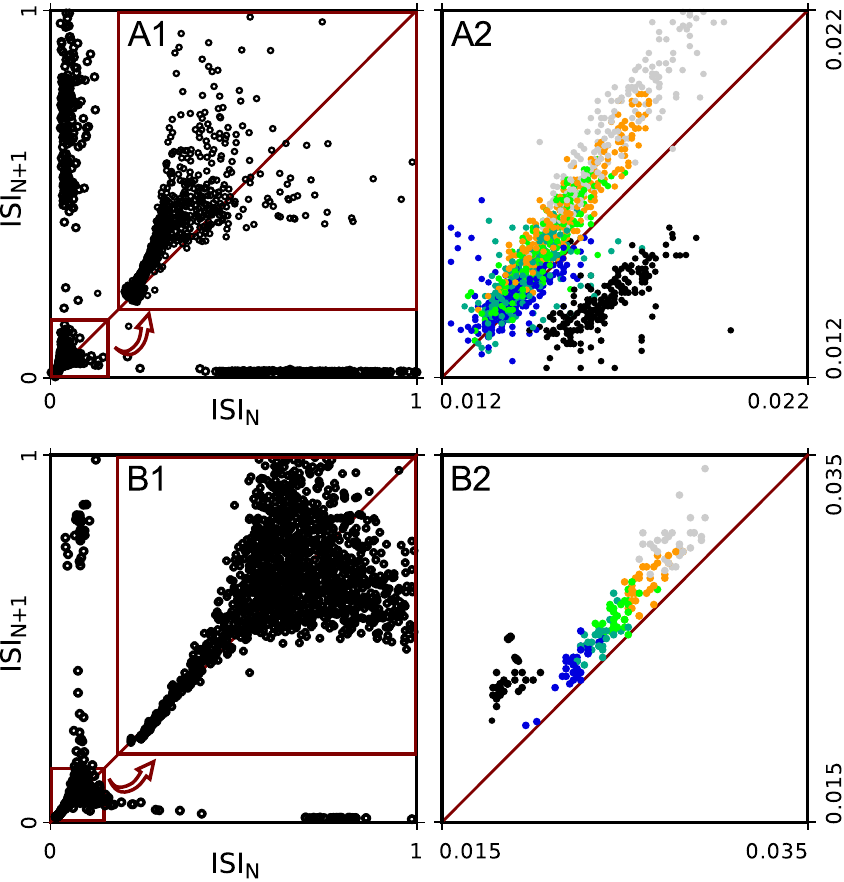}
\caption{\label{fig:expsignature}First return \isi maps for bursting
activity of isolated {\sc lp}(A) and \textsc{pd}(B) neurons from the
\stg of the lobster \emph{Panulirus interruptus}. Colours (A2, B2)
indicate successive \isi pairs at the start of bursts. \isi have been
normalized so that max(\isi) = 1.}
\end{figure}

The biological neurons, \textsc{lp} and \textsc{pd}, generated
irregular spiking-bursting activity of the type shown by the excerpted
time-series in Fig.~\ref{fig:burstlet}C. Maps of the \isi{}s for spike
trains within bursts are shown in the boxed areas of Fig.~
\ref{fig:expsignature}. The initial \isi{}s, from the start of bursts,
are shown in panels A2, B2. As the bursts evolve, the dispersion of
corresponding sequential \isi{}s increases greatly. Bursts also vary
in total number of spikes.

\subsection{Unidimensional reductions of neuronal model}

We now construct a hybrid (deterministic/stochastic) mechanism for
generating \isi map signatures similar to those of biological
neurons. It is convenient to start with unidimensional reductions of
the model (eq. \ref{eq:model}) to guide the intuition, and then
generalize to the full system. A number of different strategies for
performing such reductions have been proposed~\cite{Medvedev2005,
Medvedev2006, ChannellJr2007}, all of which could be equivalently
employed. Our analysis relied on straightforward Poincaré mapping and
fast-slow subsystem decomposition~\cite{Medvedev2005, Fenichel1979}.

Since \isi signatures are defined in terms of subsequent maxima in
membrane potential traces, the Poincaré surface of section $\Sigma$
had to be constructed in a way that the time between crossings
corresponded to intermaxima intervals for the $V$ variable. Such
requirement was satisfied by adopting the surface defined by $\dot{V}
= 0$ (see Fig. \ref{fig:phase_portrait}B for a schematic
representation of $\Sigma$ and a trajectory for the full model).

In the slow-fast decomposition, \mkm is treated as a control parameter
for the fast subsystem. We built unidimensional maps characterizing the
full dynamics by gridding the interval of \mkm values that supported
limit cycles in the fast subsystem, and using intersections of $\Sigma$
with these cycles as initial conditions for integrating the full system
(eq. \ref{eq:model}).

\begin{figure*}
\includegraphics[scale=1]{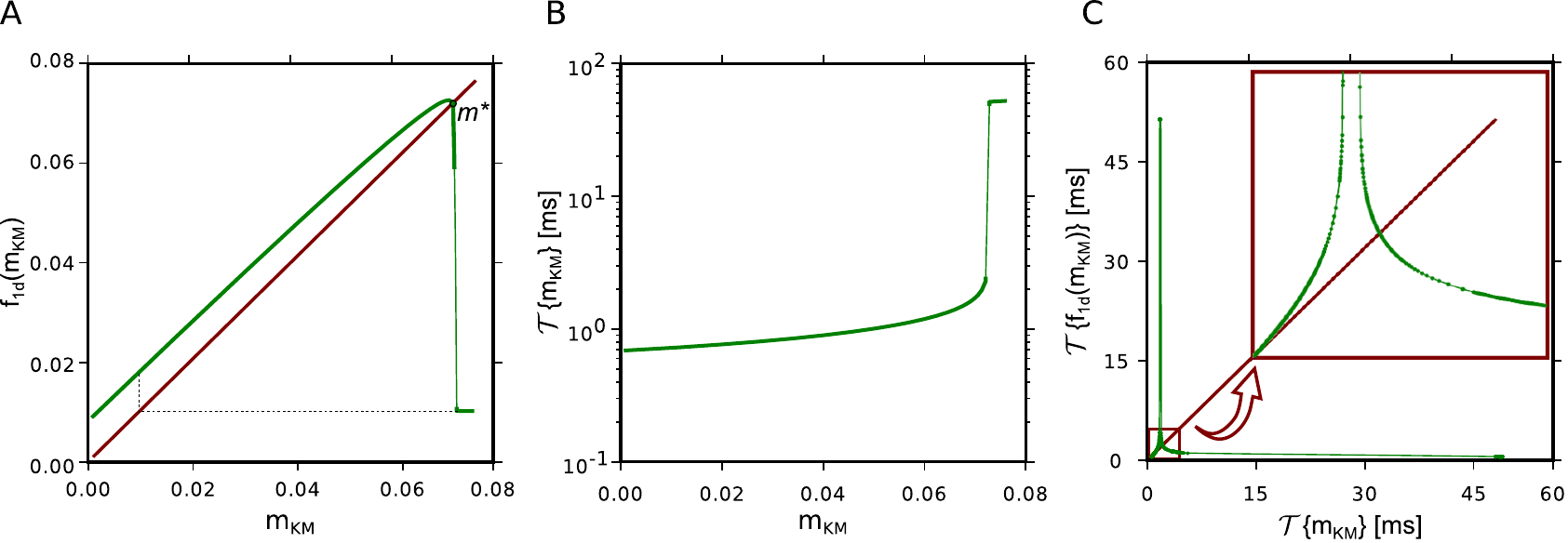}
\caption{One dimensional reduction of the dynamics. Panel A: discrete
dynamics $f_{1d}(\mkm)$ of the slow variable. The highly-negative
derivative section close to the fixed point $m^*$ is of paramount
importance in the suppression/addition of burstlets at the end of a
burst, leading to complex \isi signatures. B: the observable
\tof{\mkm} coupled to the dynamics in A, that encodes the (full model)
time elapsed in a $f_{1d}(\mkm)$ mapping. C: The \isi signature,
defined in terms of the observable \tof. The thin lines connecting
calculated points were added to guide the eye.}
\label{fig:mkm_maps}
\end{figure*}

The discrete dynamics of \mkm in the intersection of the Poincaré
surface of section $\Sigma$ with the limit cycle manifold
$\mathcal{L}$ is depicted in Fig. \ref{fig:mkm_maps}A. The mapping
$f_{1d}(\mkm)$ is the updated value of \mkm obtained by integrating
the system along a cycle starting from the initial conditions
described above, which provided us the (full model) time elapsed
between each $\Sigma \cap \mathcal{L}$ crossing. Hence, we were able
to couple an ``observable'' \tof{\mkm} to the dynamics, giving rise to
the map in Fig.~ \ref{fig:mkm_maps}B. The \isi signature map is then
straightforwardly defined in terms of this observable, as the pairs
$\left (\tof{\mkm}, \tof{f_{1d}(\mkm)} \right)$, displayed in
Fig.~\ref{fig:mkm_maps}C.

\subsection{Mechanism of noise-induced irregularity}

The apparent discontinuity in the mapping $f_{1d}$ is instrumental
in understanding irregularities in the number of spikes and total
burst duration. Notice that the dynamics is not chaotic: the strongly
dissipative quasi horizontal segment ($\mkm \approx 0.07$) reinjects
all trajectories into neighbouring points at the beginning of the
spiking manifolds $\mathcal{L}$. Nevertheless it is possible -- due to
noise -- that trajectories reach the almost vertical region of
$f_{1d}$, being mapped leftward and climbing back up the ``tube''.
Thus, the large negative derivative in $f_{1d}$ amplifies
microscopic noise, leading to irregular \isi patterns and burst
durations even when the noiseless system supports only periodic
bursting.

\begin{figure}[htbp!] 
\includegraphics[scale=1]{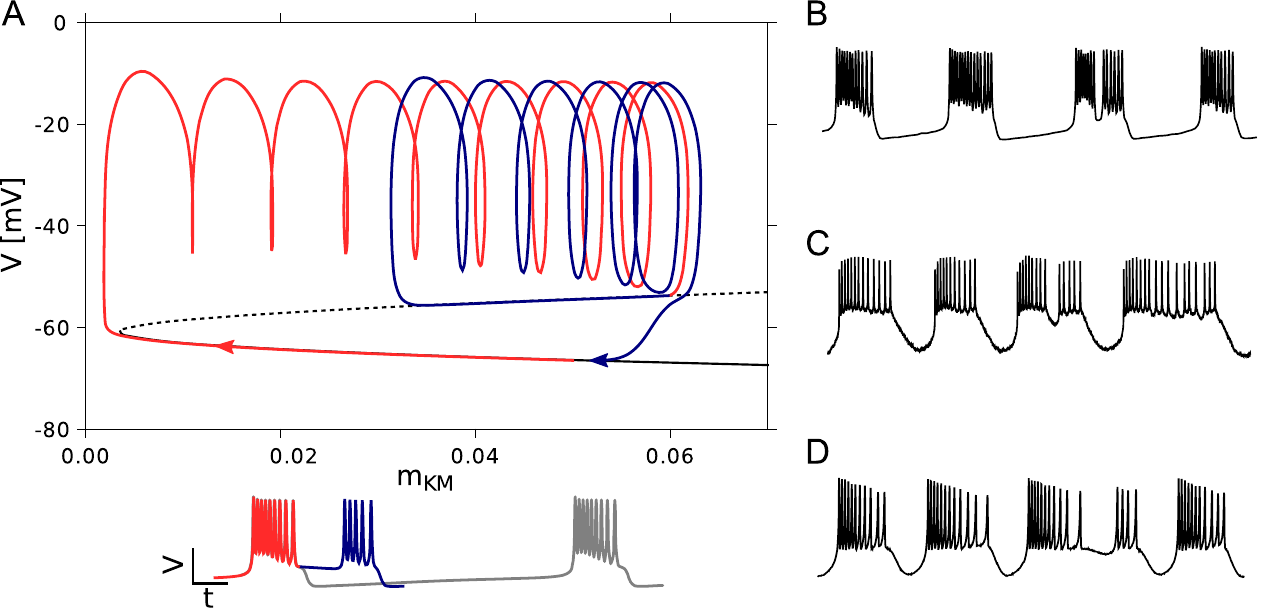}
\caption{\label{fig:burstlet}Burstlets associated with canard
trajectories.  (A): The red part of the trajectory indicates the
bursting ``main sequence'', with the flux evolving through the spiking
manifold $\mathcal{L}$. This particular trajectory does not fall
directly to the stable part of the equilibria manifold (black curve)
after $\mathcal{L}$ loses stability in the homoclinic
bifurcation. Instead, it glides along the unstable (saddle) branch of
$\mathcal{E}$ (dashed curve) for some time (blue part of trajectory),
until being reinjected into $\mathcal{L}$, thus generating a
burstlet. The bottom traces represent the timecourse of the burstlet
trajectory, superimposed with a regular (gray) one. (B, C, D):
burstlets in the stochastic \nap-\kd-\km{} model, a biological
\textsc{pd} neuron and the stochastic Hindmarsh-Rose model,
respectively.}
\end{figure}

Focusing back on the full model, it is possible to determine the
origin of the abrupt, though continuous, change in $f_{1d}$ after the
fixed point $m^*$. There is an ensemble of states close to the end of
the spiking manifold $\mathcal{L}$ that, when evolving towards
hyperpolarization, follow the saddle branch of the equilibrium
manifold $\mathcal{E}$ -- as depicted in Fig. \ref{fig:burstlet}A --
and are eventually reinjected into the spiking manifold
$\mathcal{L}$. In the membrane potential timeseries, such reinjections
would be reflected as prolongation of bursts by addition of spikes or
``burstlets'': clusters of spikes appended to a bursting trajectory,
after a hyperpolarization smaller than the typical interburst
hyperpolarization.  Examples of such burstlets can be seen in
Fig. \ref{fig:burstlet}(B,C,D).

\begin{figure}[htpb!]
\includegraphics[scale=1]{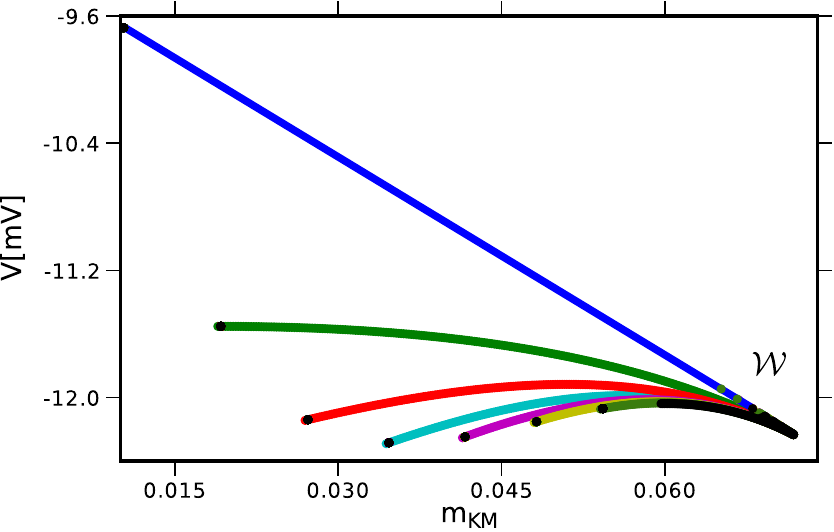}
\caption{Projection of $\Sigma$ crossings into the $\mkm V$ plane, for
initial conditions close to the branch of saddles of the fast
subsystem (segment of $\mathcal{E}$ between \textsc{sn} and
\textsc{s-h} in Fig. \ref{fig:phase_portrait}). As expected, it joins the
extremes of the 9-spike periodic orbit of the unperturbed model at the
endpoints of each curve. Other colours indicate further iterates of
the blue line. The emerging wing-shaped structure is called
$\mathcal{W}$.} \label{fig:wing}
\end{figure}

Trajectories that follow unstable structures such as the middle branch
of $\mathcal{E}$ are called \emph{canards} \cite{Rotstein2012,
Burke2012}. Note that these reinjections into $\mathcal{L}$ can take
place at any \mkm value up to the vicinity of the saddle-node
bifurcation (see Fig. \ref{fig:wing}), depending on how long the
trajectory follows the saddle branch.  It is precisely this set of
canard orbits that gives rise to the ``dynamical skeleton'' of \isi
signatures, the deterministic substrate that is populated when noise
is added to the model.

\subsection{Skeleton of the ISI signature map for the full model}

\begin{figure}[htbp!]
\includegraphics[scale=1]{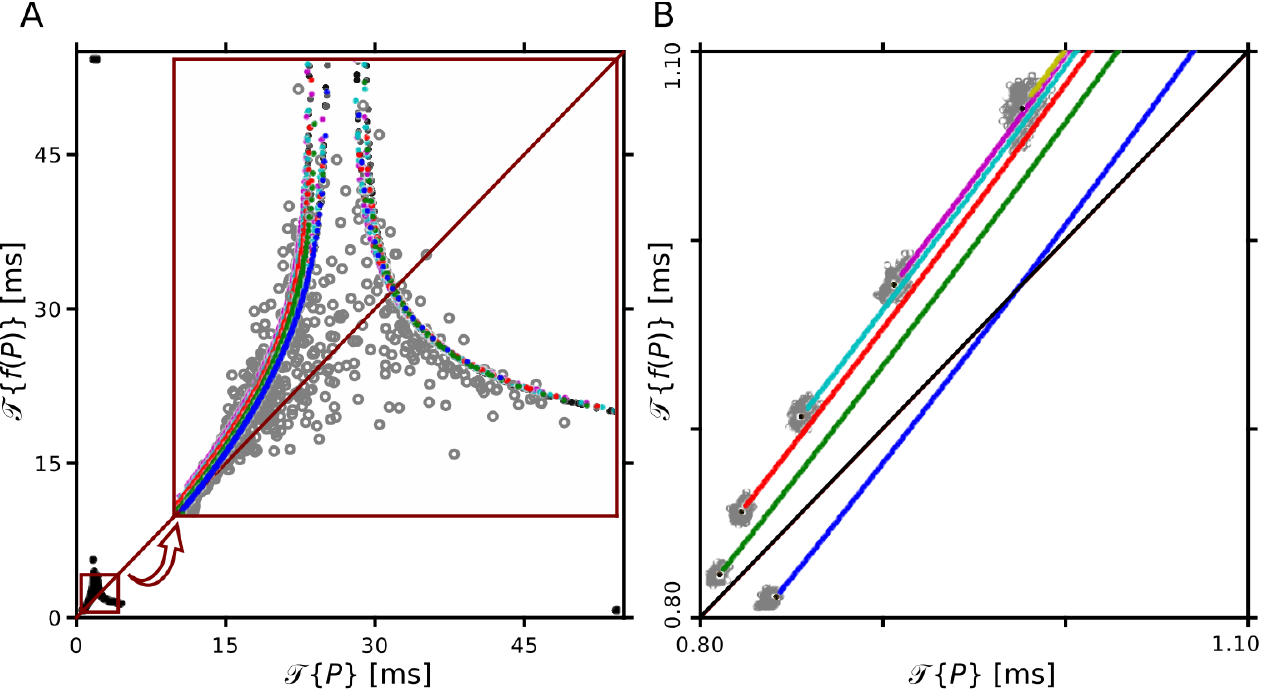}
\caption{Signature skeletons (filled circles): $\left (\ttof{P},
\ttof{f(P)} \right)$ pairs for points $P$ in the flabellate structure
$\mathcal{W}$ of Fig. \ref{fig:wing}. The generated
``infrastructure'' is only accessible by the system through the
addition of noise, given the strongly attractive character of the
periodic bursting orbit. The unfilled grey points correspond to a
signature generated by simulating the stochastic model. Panel B
depicts the cluster generation mechanism for the first \isi{}s in a
burst, along the deterministic scaffolding (coloured curves).}
\label{fig:skeleton}
\end{figure}

Let $f$ be the the discrete dynamics on $\Sigma$, i.e. it generates a
sequence of $\Sigma$ crossings according to the full model dynamics,
analogously to $f_{1d}$ for the reduced model. Let $\mathscr{T}$ be
the time elapsed between two subsequent $\Sigma$ crossings,
analogously to $\mathcal{T}$ for the unidimensional case. In order to
unearth the skeleton, we take initial conditions over the unstable
separatrices (approximated through the eigenvector corresponding to a
positive eigenvalue) of a set of saddle points spanning the middle
branch of $\mathcal{E}$, and integrate the full system until the first
$\Sigma$ crossing. These crossings give rise to the blue curve in
Fig. \ref{fig:wing}. Subsequent iterations of this curve (each
iteration is plotted with a different color in Fig. \ref{fig:wing},
and corresponds to integrating each point until the next $\Sigma$
crossing) give rise to a flabellate structure $\mathcal{W}$ in
$\Sigma$. The \isi signature skeleton is finally obtained as the
$\left ( \ttof{P}, \ttof{f(P)} \right )$ pairs for all points $P$ in
this structure, as shown in Fig. \ref{fig:skeleton}.

\begin{figure}[htbp!]
\includegraphics[scale=1]{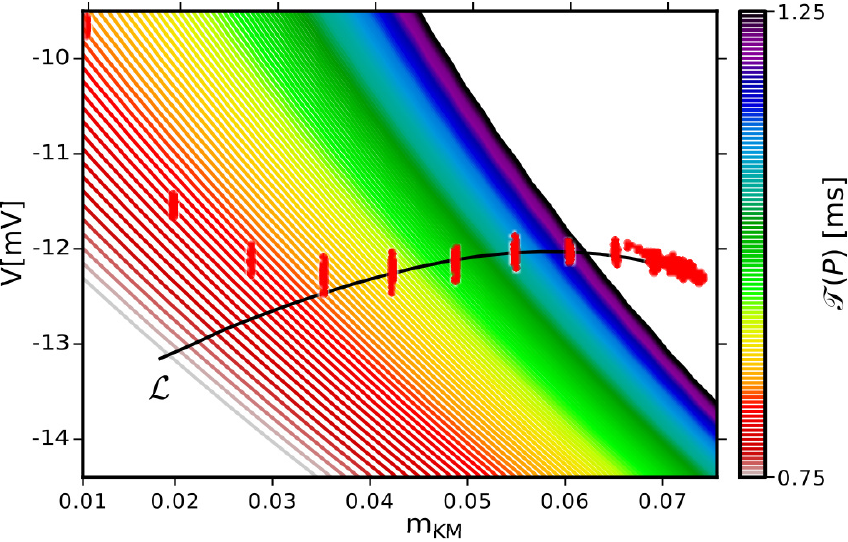}
\caption{\label{fig:isochrones}Isochrons of return time to
$\Sigma$. Long times (blank areas) have been discarded to increase
readability. The black curve is the intersection of the surface of
section $\Sigma$ with the fast subsystem limit cycle manifold
$\mathcal{L}$, corresponding to intraburst spiking.  Red dots
correspond to $\Sigma$ crossings of an integrated trajectory of the
full noisy model (equations \ref{eq:model} and
\ref{eq:langevin}). Noise tends to spread the crossing points across
isochrons, so that spikes of different positions in a burst give rise
to similar \isi{}s -- generating ``kinks'' in the signature.}
\end{figure}

Fig. \ref{fig:isochrones} represents the dynamics in $\Sigma$
projected onto the $\mkm V$ plane, with the addition of first return
isochrons. An isochron is a subset of $\Sigma$ with constant return
time. In Fig. \ref{fig:isochrones}, a colour was assigned to each
isochrone. The origin of the hook-shaped structure (``kink'') for the
smallest \isi in the experimental signatures
(Fig. \ref{fig:expsignature}-AB2, black-blue-green sequence) is
elucidated by noticing that noise tends to spread crossings across
isochrons.  This way, two \isi in different positions along the burst
are similar, leading to vertically stacked clusters in the $\left
(\ttof{P}, \ttof{f(P)} \right)$ (signature) map.

\subsection{Simulation of \isi map signatures}

\begin{figure}[htp!]
\includegraphics[scale=1]{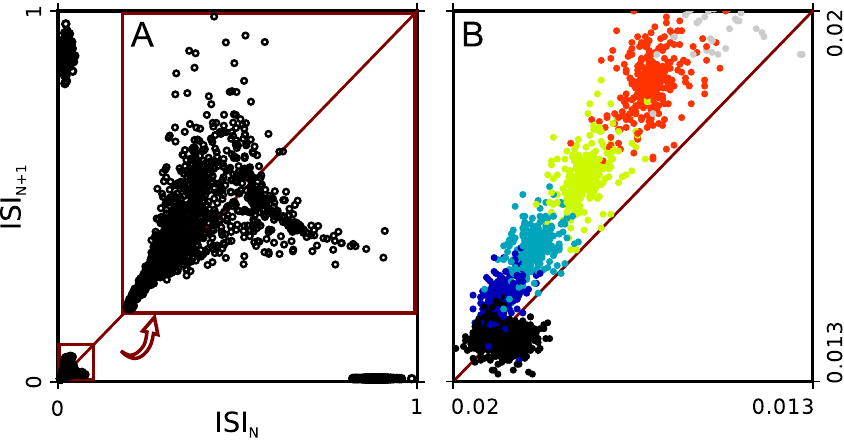}
\caption{\isi signature generated by the model of a square-wave burster
neuron with channel noise (eqs. \ref{eq:model} and
\ref{eq:langevin}). Scaling of plots and point colors as in
Fig. \ref{fig:expsignature}.}
\label{fig:modelsignature}
\end{figure}

\begin{figure}[htp!]
\includegraphics[scale=1]{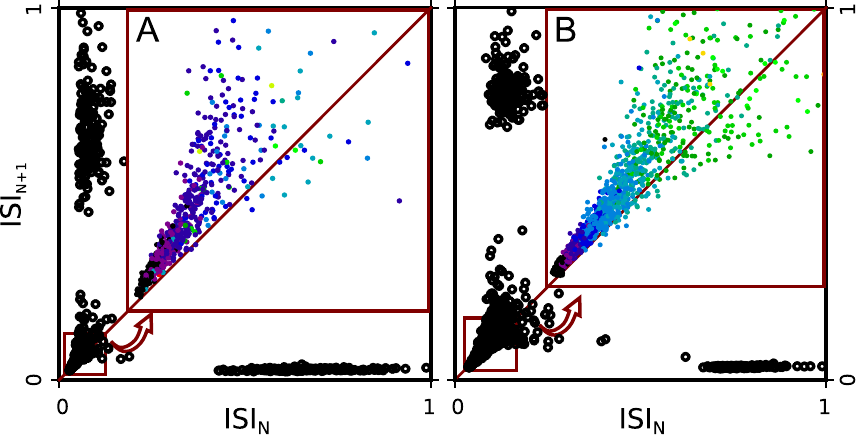}
\caption{(left): Signature generated by the pre-Bötzinger neuron model
with conductance noise; (right): Signature generated by the
Hindmarsh-Rose neuron model with current noise. Conventions as in
Figs. \ref{fig:expsignature} and
\ref{fig:modelsignature}.} \label{fig:hrbotzinger}
\end{figure}

The integration of the full (noisy) model (eqs. \ref{eq:model} and
\ref{eq:langevin}), in addition to a simple threshold ($-30\text{mV}$)
spike detection, leads to the \isi signature map in Fig.
\ref{fig:modelsignature}.  This map qualitatively reproduces the
structure of the \isi signature of biological neurons
(Fig. \ref{fig:expsignature}), including fine details such as the
``kink'' in the low \isi main sequence, as well as the variability in
both number of spikes and interburst intervals (reflected in the
diffusiveness at the end of the main sequence and the extent of the
isolated high-\isi arms).  All of these features can be traced back to
an underlying deterministic scaffolding, so that the role of noise is
to induce transient dynamics exposing this infrastructure.  Note also
that the qualitative resemblance between model and biological \isi maps
was obtained without special tuning of model parameters. Similar
structures were also obtained by adding noise to either pre-Bötzinger
or Hindmarsh-Rose model neurons (Fig. \ref{fig:hrbotzinger}),
suggesting that similar mechanisms may be at work in these cases also.

\section{Discussion}

The presented mechanism accounts for the generation of irregular
bursting traces with complex signatures, in terms of a low-dimensional
conductance-based model and a macroscopic approximation of stochastic
gating noise. Noise plays a crucial role in the mechanism: although
the deterministic scaffolding of the model (its ``skeleton'') can
support complex behaviour, this dynamical richness is usually
suppressed by the dissipative character of the periodic bursting or
tonic spiking orbits. Noise, however, unveils the transient dynamics,
giving flesh to the skeleton and generating the \isi signature
patterns characteristic of biological neurons. Let $\Sigma$ be the
hypersurface defined by $\dot{V} = 0$, which will be [n-1]-dimensional
in a n-dimensional conductance-based model.  The set of intersections
of $\Sigma$ with the unstable ([n-1]d) manifolds of the saddles in the
middle branch ([n-2]d) and its iterates defines the flabellate
structure $\mathcal{W}$. Skeletons are the image of the \isi return
map transformation applied to $\mathcal{W}$.

We emphasize the robustness of the skeleton to parameter fluctuations:
since it is inherently tied to the bifurcation structure of the model,
its general features persist even through bursting-tonic transitions
(associated with the gain of stability of the fixed point in the
unidimensional map $f_{1d}$).  Signatures essentially will remain
the same for scenarios in which bursting is induced by noise (the
noiseless system otherwise spiking tonically)~
\cite{Hitczenko2007}. That seems indeed to be the case for the
\textsc{pd} neuron (Fig.~ \ref{fig:expsignature}B), given the presence
of very long (more than 100 spikes) bursts and comparatively shorter
hyperpolarization periods.

Irregular activity in neuronal models has been associated with the
presence of deterministic
chaos~\cite{Falcke2000,Carelli2005,Channell2009}. Nevertheless, the
main cause of irregularities in our model is the amplification of
stochastic phenomena by the transient dynamics. Structures defined by
such dynamics persist even for parameter regimes that do not support
chaotic attractors. Noise-induced chaos~\cite{Liu2002} -- where the
neighbourhood of non-attracting hyperbolic sets is visited due to
perturbations-- could be present, as chaotic saddles can arise in
spike-adding transitions as shown in~\cite{Terman1991}.  This would,
however, involve fine tuning of parameters and might prove too
delicate to detect~\cite{Gao1999} with large noise intensities such as
those needed, seemingly, to simulate biological results. Similar
considerations may apply also to ``stochastic chaos'' associated
with D-type stochastic bifurcations~\cite{Kosmidis2006}.

Bursting activity can be generated through several distinct geometric
mechanisms~\cite{Izhikevich2000} in addition to the
saddle-node/homoclinic behavior studied here. Using the proposed
geometrical framework, general features of the dispersion of
\textsc{isi} pairs in neuronal signature maps can be predicted. In
particular, the burstlet generation mechanism via canard trajectories
will require bursting scenarios involving a saddle middle branch.

As different levels of noise are added to the model, there is a
scaling of \isi cluster dispersion (data not shown). This provides a
possible method of estimating the dynamical noise level in time series
analysis of real neurons.  We also point out that the burstlet
definition and description can be used to improve burst alignment
algorithms~\cite{Lago-Fernandez2009}, through separating bursts into a
``main sequence'' followed by irregular burstlets.

\begin{acknowledgments}
Financial support from the Brazilian agencies \emph{Fundação de Amparo
à Pesquisa do Estado de São Paulo} (\textsc{fapesp}), \emph{Coordenação
de Aperfeiçoamento de Pessoal de Nível Superior} (\textsc{capes}) and
\emph{Conselho Nacional de Desenvolvimento Científico e
Tecnológico} (\textsc{cnpq}) is gratefully acknowledged. \textsc{rce}
was supported by a grant from \emph{National Science Foundation}. We thank
Geoffrey Evans for useful feedback on the manuscript.
\end{acknowledgments}

\bibliography{references}

\end{document}